# Internal Stresses and Formation of Switchable Nanowires
# at Thin Silica Film Edges

J. C. Phillips

Dept. of Physics and Astronomy, Rutgers University, Piscataway, N. J., 08854-8019

## INTRODUCTION

At vertical edges, thin films of silicon oxide ($SiO_{2-x}$) contain semiconductive c-Si layered nanocrystals (Si NC) embedded in and supported by an insulating g-$SiO_2$ matrix. Tour *et al.* have shown that a trenched thin film geometry enables the NC to form switchable nanowires (SNW) when trained by an applied field. The field required to form SNW decreases rapidly within a few cycles, or by annealing at 600 C in even fewer cycles, and is stable to 700C. Here we describe the intrinsic evolution of Si NC and SNW in terms of the competition between internal stresses and electro-osmosis. The analysis relies heavily on experimental data from a wide range of thin film studies, and it explains why a vertical edge across the planar Si-$SiO_x$ interface is necessary to form SNW. The discussion also shows that the formation mechanisms of Si NC and Si/$SiO_{2-x}$ SNW are intrinsic and result from optimization of nanowire conductivity in the presence of residual host misfit stresses.

Many semiconductive materials (such as metal oxides and organic compounds) have exhibited resistive switching, and the phenomenon is appealing for random access memory (RAM) applications. However, the phenomenon has usually been non-reproducibly associated with grain boundaries or electrode interfaces[1]. It has often been assumed that it is exponentially difficult to form native filaments (as distinguished from metallic filaments extended from metallic electrodes) in semiconductive materials. Native filaments can be formed electro-osmotically[2] in Te-based chalcogenide alloys[3] because Te is ambi(2 or 3)-valent, but the switching apparently occurs at electrodes[4].  The presence of metallic filaments formed intrinsically by electron-electron interactions explains[5] the intrinsic power-law metallic conductivities with $\alpha = 0.50 \pm 0.01$ of low-temperature insulator-metal transitions in some impurity band semiconductors even with randomly distributed impurities (Si:P), but these networks are not switchable.

In terms of complementary metal oxide semiconductor (CMOS) technology, by far the most appealing material is $SiO_x$ itself, which can be a mixture of semiconductive Si 5 nm nanocrystals (NC) embedded in an insulating $g-SiO_2$ matrix.  However, it was long supposed that the switching behavior sometimes reported for $SiO_x$ mixtures was associated with conduction through metal filaments from the electrodes[6].  A series of remarkable experiments[6] has shown intrinsic formation of switchable nanowires (SNW), provided that an apparently irrelevant geometry is involved: there must be a vertical edge across the planar substrate Si-$SiO_x$ interface, which normally has corners where the silica edges meet the Si substrate.  Here we show that this geometrical condition can be explained by the absence of long-range elastic forces involving shear induced at the planar Si-$SiO_x$ interface by SNW formation when the corners are removed and replaced by a vertical edge. We also discuss evidence that indicates that the Si 5 nm NC are intrinsic to the host edge environment. The positions and orientations of the NC are metastabilized by inhomogeneous internal network stresses, which are known to be large in rigid, refractory network materials like Si, and a "hidden" driver of network self-organization[7].

As we shall see, the formation of SNW, and even the emergence of Si NC, near thin film $SiO_{2-x}$ edges is made possible by a delicate balance between long-range elastic and electro-osmotic forces.  This balance cannot be evaluated quantitatively because it contains too many spatially variable factors that are inaccessible to experimental observation or numerical simulation. However, these factors are also present in many other experiments, especially ones with either similar geometries, with similar stresses, but different materials, or with similar Si/$SiO_{2-x}$ chemistries and different geometries.  Analysis of these complementary experiments enables us to rescale what is known about sub nm molecular Si/$SiO_{2-x}$ chemistry to the 5-200 nm length scale of SNW formation.

# RELATED EXPERIMENTS

The first questions involve the Si 5 nm NC observed at the edge in Fig. 3 III of ref. 6, as one would not have expected to find such perfect Si NC in a glassy $SiO_{2-x}$ environment with x ≪ 1. Layers of a-Si (thickness ~ 10 nm) alternating with thin (thickness ~ 5 nm) $SiO_2$ layers can be crystallized either thermally[8] or by laser annealing[9], with Si NC diameters shown in Figs. 1 and 2. The thermal Si NC are perfect up to 5 nm diameter, and contain defects (such as dislocations) for larger diameters[8]. Careful inspection of Fig. 1 suggests that a better fit to the data is given with two lines, rather than the authors' smooth curve, with the break in slope coming near 5.0(2) nm diameter. The smaller and perfect NC are placed[10] near the center of the 10 nm a-Si space between the $SiO_2$ confining walls, while the filling factor slope for the larger imperfect NC is smaller.

Laser-annealed Si NC also exhibit evidence for intrinsic perfect Si NC with diameters $s_{nc}$ up to 5 nm. As seen in Fig. 2, in 10 nm a-Si/3 nm $SiO_{2-x}$ MQW, Si NC diameters (estimated from FWHM Raman bands) grow with increasing laser power $P_{LAS}$ in two stages, first nearly isotropically (presumably nucleating at the centers of the a-Si layers[10]), with maximum $ds_{nc}/dP_{LAS}$ near $s_{nc}$ ~ 5 nm, then flattening below the MQW width of 10 nm, and then exhibiting a second maximum near $s_{nc}$ ~ 10 nm, suggestive of formation of granular 2x2 Si NC plaquettes formed from perfect 5 nm Si NC.

Si NC strains $\varepsilon_{MQW}$ are also shown in Fig. 2, with positive $\varepsilon_{MQW}$ being compressive, and $\varepsilon_{MQW}$ reversing sign near $s_{nc}$ = 5 nm. This sign reversal was attributed[9] to the smaller MQW filling factor of c-Si relative to a-Si, but as the Si NC are only weakly stressed by the much more plastic a-Si, it appears that the sign reversal could also be caused by the appearance of imperfections (such as dislocations) in the NC for $s_{nc}$ > 5 nm. At $s_{nc}$ = 5 nm, Si NC are rigid but unstressed, much like isobaric network glasses[7], which favors their stress-free matching to the glassy $SiO_{2-x}$ host film planar boundaries mediated by a-Si.

Why is $\varepsilon_{MQW}$ compressive for small $s_{nc}$? This is a topological result which is connected to the compacting effects of back-bonding in small Si clusters; the back bonds are shortened by interactions with dangling bonds[11,12]. Enhancement of back-bonding interactions in a classical force field (CFF) gave an excellent description of small and medium size ($n < 30$) $Si_n$ cluster structures[11]. In incompletely oxidized $SiO_{2-x}$ films ($0 < x \ll 1$) Si atoms can segregate near free edges to exploit back-bonding and form clusters with $s_{nc}$ as large as, or larger, than 0.4 nm[12]. Although it lies outside the scope of the present work, this CFF, together with an $SiO_2$ CFF, could be used in an MDS to examine Si NC and SNW growth near free edges.

## SNW FORMATION

Perfect Si NC may be present at the edges of as-deposited planar 40 nm $SiO_{2-x}$ ($0 < x < 0.1$) film, but the low-resistive state is formed initially only after an application of V > 20V. For a large enough voltage, current paths centered on ~ 5 nm Si nanocrystals (NC) can be activated, but initially these paths will have a complex topology, with their segments broken and misoriented. The NC can be re-oriented to form more nearly vertical continuous nanowires by oxygen vacancy diffusive flow around NC faces and into the broken regions. However, such reconstruction to enhance the electrical conductivity $\sigma$ increases the as-grown internal misfit stresses. Because the compressibility is much larger than the shear modulus, SNW formation induces shear and can even destroy Si NC, especially at the edge of the soft and adaptive planar Si-$SiO_x$ interface. The latter is one of the specially selected alloy compositions which is unstressed at the molecular level, and so can have defect (trap) densities of order $10^{-4}$ with a density difference of 30%[13]. External evidence for shear-induced deformations of vertical edge interfaces is shown in Fig. 2(c) of ref. 6.

As shown in Fig. 3 III of ref. 6, the Si NCs align electo-osmotically in a pathway that is parallel to the current direction (the switching path), with (111) surfaces in the edge plane. The Franz-Keldysh electric field-induced electronic states associated with a quasi-one-dimensional filamentary state, or a quasi-two-dimensional edge state, are enhanced by the lowered dimensionalities relative to those for three-dimensional geometries[14], microscopically facilitating classical electro-osmosis. While the NC's are formed and connected at large V ~ 20V, after training the limiting switching voltage for 40 nm thick films is $V_{LIM}$ = 3.8 V. Experiments have shown[15] that both $V_{LIM}$ = 3. 8 V and its 0.3 V spread are independent of film thickness t for 40 nm < t < 200 nm. The limiting voltage may be associated with a transition state associated with excitation of a Si bonding electron to the Fermi energy of the silica host. The observed value of $V_{LIM}$ is somewhat smaller than the difference in silica of 4.36 eV between the Fermi energy of silica and the VBM of silicon (111) in the silica carrier regime[16], because of the gain in energy associated with network crystallization in the gaps between Si NC.

The shear stress of the interface provides a large mechanical energy barrier energy for SNW reconstruction when it is associated with a sharp corner at the film edge, where the residual elastic stresses are nearly singular. The significance of corners in place of vertical edges has been demonstrated previously by shear-induced vertical ordering of ferroelectric nanocrystals in spin-coated thin film polymers[17] and by disappearance of stress singularities at interface edges in vertically nanostructured thin film columnar oxides[18]. It has also been calculated by a finite element analysis for two types of components with different interface edges between the thin film and substrate[18] (see Fig. 3). There are also examples for metallic[19] (blocking effect of notches) and colloidal[20] (shear disruption of a conductive network) glasses, and model calculations of hysteresis loops in rugged potential landscapes[21]. The dynamics of shear-related interfacial effects have also been studied[22] in electrorheological fluids, which are suspensions of particles having higher dielectric constant or electric conductivity than that of suspending fluids with a low viscosity (similar to Si nc embedded in $SiO_x$ glass films). When the particles have permanent dipole moments, electric fields produce giant effects, with the formation of easily observed ferroelectric channels[23]. When Si NC nanowires are formed near vertical $SiO_x$ edges, resistive switching voltages are independent of film diameter, a desirable device property[6].

## CONCLUSIONS

The remarkable inducible SNW properties reported for $SiO_x$ mixtures are intrinsic, and are made possible by the stability and elastic properties of 5 nm Si NC, and the absence of singular sheer stress SNW rupture at planar Si-SiO$_x$ interfacial corners.  Given the rapid evolution of nanoscale science, it may be possible to study these divergent stresses even at the monolayer level, for instance in ceramic high-temperature superconductors, which also contain self-organized conductive nanowires[24].

## Figure Captions

Fig. 1.  Si NC sizes grown in thermally annealed a-Si sublayers confined between

$SiO_{2-x}$ layers.  The figure is from ref. 7, and the smooth curve has been supplemented by two line segments to emphasize the effects of NC perfection on $s_{nc}$.

Fig. 2. Growth of Si NC from a-Si confined to a 10 nm layer between $SiO_{2-x}$ layers, from ref. 8.  Arrows have been added to mark fastest growth rates (see text).

Fig. 3.  The calculations of ref. 18, illustrated here with their Fig. 8 (shown here for the reader's convenience), treat stress fields near (a) free edges (which support SNW) and (b) corners (which do not support SNW), where cracks can form.  The theory of ref. 18 is accompanied by experiments on samples where the Si substrate supports a thin film composed of $Ta_2O_5$ helical nanosprings, and confirm the singular character of stress fields near nanocorners.  Note that in the Si NC/$SiO_{2-x}$ case, one could have a disruptivechemical crack in the Si NC concentration at a corner with the Si substrate without having a crack at the physical $SiO_{2-x}$ edge/Si substrate corner itself.

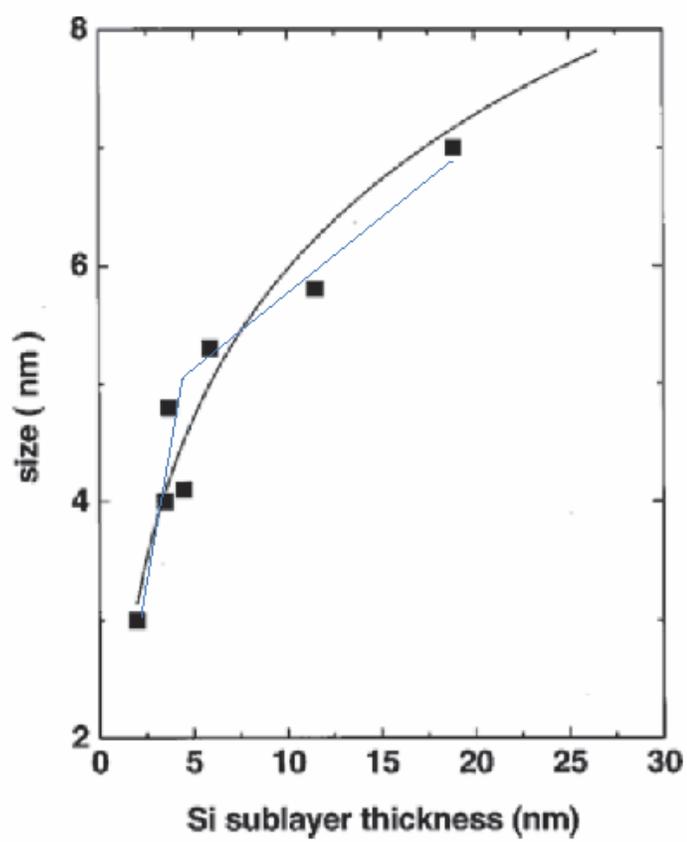

Fig. 1.

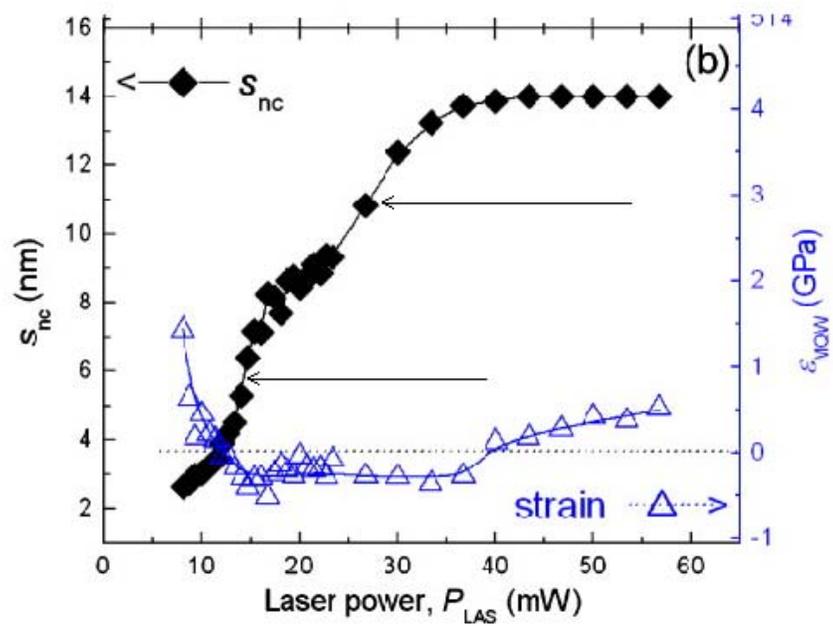

Fig. 2.

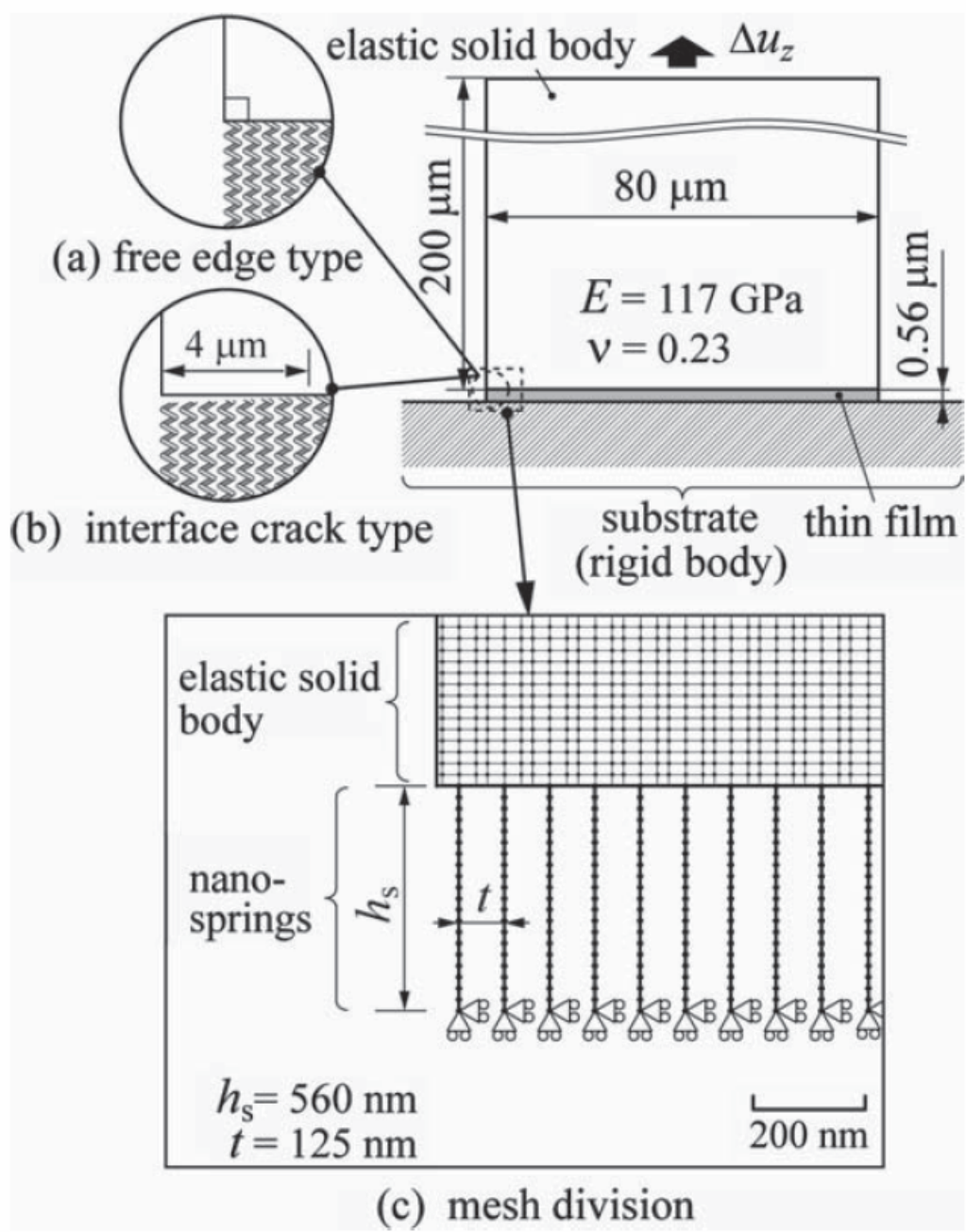

Fig. 3.